\documentclass[prb,twocolumn]{revtex4}

\usepackage[dvips]{graphicx}
\usepackage{latexsym,amsmath,amssymb,bm,euscript}
\usepackage[dvips]{color}

\begin{document}

\title{Line of continuous phase transitions in a three-dimensional U(1) loop model with $1/r^2$ current-current interactions.}
\date{\today}

\author{Scott D. Geraedts}
\author{Olexei I. Motrunich}
\affiliation{Department of Physics, California Institute of Technology, Pasadena, California 91125, USA}

\begin{abstract}
We study a lattice model of interacting loops in three dimensions with a $1/r^2$ interaction. Using Monte Carlo, we find that the phase diagram contains a line of second-order phase transitions between a phase where the loops are gapped and a phase where they proliferate. The correlation length exponent and critical conductivity vary continuously along this line. Our model is exactly self-dual at a special point on the critical line, which allows us to calculate the critical conductivity exactly at this point.
\end{abstract}
\maketitle

\section{Introduction}
Continuous phase transitions form a fascinating subject in statistical mechanics.\cite{DombGreen}  They are understood essentially completely in two dimensions\cite{Baxter, CFT} and can be described in mean field in high enough dimensions.  For intermediate dimensions, much knowledge is obtained from field-theoretic treatments\cite{Ma, Cardy} such as large-$N$ and $4-\epsilon$ expansions, as well as from numerical Monte Carlo simulations.\cite{MCBook}  In this paper, we consider a class of statistical mechanics models with a global U(1) symmetry and specific \emph{marginally-long-ranged} current-current interactions\cite{Fradkin_SL2Z, Kuklov2005} that decay as $g/r^2$ in three dimensions (3d), where $g$ is the coupling.  Our Monte Carlo study suggests that these have continuous phase transitions with critical properties such as correlation length exponent and critical conductivity that vary as a function of the coupling $g$.  Our lattice model is self-dual for a special value of the coupling and we know the location of the continuous phase transition exactly, a rare instance in 3d statistical mechanics problems. This means that we know the critical conductivity exactly, but not the correlation length exponent, for which we need the Monte Carlo simulations performed in this paper.

One of the motivations for our study is the problem of phase transitions in matter-gauge systems,\cite{HLM1974, Chen1978, ColemanWeinberg1973, Dasgupta1981, KamalMurthy1993, deccp, shortlight, nccp1new, Kuklov08,Chen,Charrier} with a schematic action
\begin{eqnarray*}
&&S = S_{\rm matter} + S_{\rm gauge} ~, \\
&& S_{\rm matter} = \int d\vec{r} \left[ |(\vec{\nabla} - i \vec{a}){\bm \Psi}|^2 + m |{\bm \Psi}|^2 + u |{\bm \Psi}|^4 \right]  ~, \\
&& S_{\rm gauge} = \int d\vec{k} ~ \frac{1}{2} \Pi(k) ~ 
\left(\delta_{\mu\nu} - \frac{k_\mu k_\nu}{k^2} \right) ~ a^*_\mu(k) a_\nu(k) ~.
\end{eqnarray*}
For concreteness, we took bosonic matter fields and also wrote the action for the gauge field in $k$-space.  Here we will consider singular gauge action with $\Pi(k) \sim |k|/g$ on long wave-lengths in 3d.  This would arise, for example, if we had a more microscopic model with additional critical matter fields\cite{Irkhin1996, RantnerWen2002, Vafek2002, Franz2003, Borokhov2002, HandsKogut2004, Hermele2004, KaulSachdev2008} that we managed to integrate out (e.g., singular gauge field propagators arise in formal large-$N$ treatments and the effective coupling $g$ depends on the number of flavors).  In the present work, we simply postulate the singular action at the microscopic level and allow ourselves to vary the coupling $g$ at will.  We also focus on the case of one U(1) matter field. For positive $m$ the field ${\bm \Psi}$ is gapped, while for negative $m$ the field condenses, and we are interested in the properties of this phase transition.

A convenient representation of the U(1) matter field is in terms of integer-valued conserved currents on a lattice, with the action
\begin{eqnarray}
S_{\rm matter} &=& \frac{1}{2} \sum_{r,r'} V_{\rm s.r.}(r-r') \vec{J}(r) \cdot \vec{J}(r') \label{matter} \\
& +& i \sum_r \vec{J}(r) \cdot \vec{a}(r) ~, \nonumber
\end{eqnarray}
where $V_{\rm s.r.}(r-r')$ is some short-range interaction. Upon integrating out the gauge field, we obtain the action in terms of current loops only, with long-range current-current interactions.

A precise definition of a general action for a system of loops with long-range interactions is:
\begin{equation}
S[\vec{J}]=\frac{1}{2}\sum_{r,r',\mu} V(r-r')J_{\mu}(r)J_{\mu}(r').
\label{general}
\end{equation}
Here $\vec{J}$ are integer-valued currents which reside on the links of a cubic lattice with periodic boundary conditions; $r,r'$ are the sites of this lattice, and $J_{\mu}(r)$ is on the link between $r$ and $r+\hat{\mu}$. The currents are subject to the constraint $\vec{\nabla} \cdot \vec{J}(r)=0$ for all $r$. 
The matter-gauge system with the matter action given in Eq.~(\ref{matter}), upon integrating out the gauge field, gives the loop action Eq.~(\ref{general}) with the potential $V(r-r')=V_{\rm l.r.}(r-r') + V_{\rm s.r.}(r-r')$, with $V_{\rm l.r.}(k) = 1/\Pi(k) \sim g/|k|$ in $k$-space and $V_{\rm l.r.}(r-r') \sim g/|r-r'|^2$ in real space. From now on, we will consider such loop-only statistical mechanics models. When the overall repulsion in Eq.~(\ref{general}) is large, the system will be in a phase with only small loops, while at small repulsion the system will be in a phase where the loops proliferate. 

\section{Model and Measurements}

Before we proceed with the direct Monte Carlo study, we first review some results from a duality approach to such current loop systems for general $V$, and will then specialize to our precise model. Consider the action
\begin{equation}
S_{\rm orig}[\vec{J}] = \frac{1}{2} \sum_k V(k) |\vec{J}(k)|^2 +i\sum_k \vec{J}^*(k)\cdot\vec{A}_{{\rm ext}}(k)~.
\label{SJ}
\end{equation}
This is the model defined in Eq.~(\ref{general}) written in $k$-space, except that it also includes a static external field $\vec{A}_{{\rm ext}}$. The partition sum $Z[\vec{A}_{\rm ext}]$ can be used to extract current correlation functions. We can use a duality transform 
\cite{Peskin1978, Savit1980, Dasgupta1981, PolyakovBook, FisherLee} to express the partition sum in terms of dual variables $\vec{Q}(R)$, which are defined on a lattice dual to the original lattice. The $\vec{Q}(R)$ are also conserved integer-valued currents, and can be viewed as vortex loops.\cite{Peskin1978, Savit1980, Dasgupta1981, PolyakovBook, FisherLee}  The action becomes
\begin{eqnarray}
&&S_{\rm dual}[\vec{Q}] = \frac{1}{2} \sum_k V_{{\rm dual}}(k) \left|\vec{Q}(k)+[\vec{\nabla} \times \vec{A}_{{\rm ext}}](k) \right|^2,
\label{SQ}\\
&&V_{{\rm dual}}(k)=\frac{(2\pi)^2}{V(k)|\vec{f}(k)|^2},
\label{duality}
\end{eqnarray} 
where $f_\mu(k) \equiv 1 - e^{i k_\mu}$, $|\vec{f}(k)|^2=\sum_{\mu}[2-2\cos(k_{\mu})]\approx k^2$ for small $k$. This duality is {\it exact} for a finite lattice if we require $\vec{J}_{{\rm tot}}\equiv \sum_r\vec{J}(r)=0$, $\vec{Q}_{{\rm tot}}=0$, which is helpful when we characterize response functions and also when we study an exact self-dual point below. 

We monitor loop behavior in our simulations by measuring the ``superfluid stiffness'' of the loops, which is defined as:
\begin{equation}
\rho^{\mu\mu}(k)\equiv\frac{1}{{\rm Vol}}\left\langle \left| \sum_{r} J_{\mu}(r) e^{i\vec{k} \cdot \vec{r}}\right|^2\right\rangle=
\left\langle|J_{\mu}(k)|^2\right\rangle~,
\label{rhoE}
\end{equation}
where Vol $\equiv L^3$ is the total number of sites. Because of the vanishing total current, we measure these at the smallest non-zero $k$. For example, for $\rho^{xx}$ we used $k_{{\rm min}}\equiv(0,0,\frac{2\pi}{L})$. We can obtain current-current correlations by differentiating the generating function $Z[\vec{A}_{\rm ext}]$ expressed either in terms of the original variables or the dual variables. Equating the two leads to relations like:\cite{HoveSudbo2000, Herzog2007}
\begin{equation}
V(k)\rho^{xx}_J(k)+V_{{\rm dual}}(k)\rho_Q^{yy}(k)=1~,
\label{Vrelation}
\end{equation}
for $k\equiv(0,0,k_z)$ which is also assumed in the formulae below. Once again, this relation is exact for a finite lattice as long as there are no total current circulations. 

The superfluid stiffness characterizes the current response of the system to an externally applied field $\vec{A}_{{\rm ext}}$.\cite{Cha1991}  However, in a system with long-range interactions there is an additional ``internal'' field created by the system's response to $\vec{A}_{{\rm ext}}$. Thus, if we start with $S_{{\rm matter}}$, Eq.~(\ref{matter}), with fluctuating internal gauge field $\vec{a}$ governed by the action $S_{{\rm gauge}}$ and add the probe field $\vec{A}_{{\rm ext}}$, we induce non-zero $\langle \vec{a} \rangle$. In order to measure the system's response to the total field,\cite{Herzog2007, MurthyShankarRMP} we define a new observable:
\begin{equation}
\sigma^{xx}_J(k)\equiv\frac{\rho^{xx}_J(k)}{|\vec{f}(k)|[1-V(k)\rho_J^{xx}(k)]}~.
\label{sigmadef}
\end{equation}
$\sigma^{xx}(0,0,k_z)$ is the Matsubara conductivity\cite{Cha1991, Damle1997} at the imaginary frequency $ik_z$. We can use Eq.~(\ref{Vrelation}) to show that the conductivities in the original and dual variables are related by 
\begin{equation}
\sigma^{xx}_J(k)\sigma_Q^{yy}(k)=\frac{1}{(2\pi)^2}.
\label{sigmarelation}
\end{equation}
In an isotropic system, $\sigma^{xx}(k)=\sigma^{yy}(k)$. From now on, we drop Cartesian indices on $\rho$ and $\sigma$.

We now consider the behavior of $\rho$ and $\sigma$ in the different phases of the model. In the small loops phase, Eq.~(\ref{rhoE}) gives $\rho_J(k) \sim k_{z}^2$, and we can see from Eq.~(\ref{sigmadef}) that $\sigma_J(k)\sim k_{z}$, and so for $k_z=2\pi/L$, both go to zero in the thermodynamic limit. In the proliferated phase, we know that the dual variables $\vec{Q}$ are in the small loops phase, and we can use the Eq.~(\ref{Vrelation}) to show that $1-V(k_{{\rm min}})\rho_J(k_{{\rm min}})  \sim V_{{\rm dual}}(k_{{\rm min}})k_{z}^2$, and Eq.~(\ref{sigmarelation}) to show $\sigma_J(k_{{\rm min}}) \sim 1/k_{z}$. 

We now specialize to our model, where we take
\begin{equation}
V(k)=\frac{2\pi g}{|\vec{f}(k)|}+t~
\label{V}
\end{equation}
for all $k=(k_x,k_y,k_z)$; $g$ and $t$ are the parameters of the model, with $g$ the strength of the $1/r^2$ interaction and $t$ the strength of an on-site interaction. This model was considered by Ref.~\onlinecite{Kuklov2005}, while our study provides a good physical characterization and understanding utilizing the connection to matter-gauge systems. 

Applying the above analysis, we see that in this model $V(k_{\rm min})\rho(k_{\rm min}) \sim 1/L$ in the small loops phase and $\sim 1-\alpha/L$ in the proliferated phase, where $\alpha$ is some non-universal number, and we see that we can use $V(k_{\rm min})\rho(k_{\rm min})$ in a manner similar to Binder ratios in magnetic systems. Since $V(k_{\rm min}) \sim L$, we also see that $\rho \cdot L$ approaches a finite number in the proliferated phase. 

In our simulations, we will keep $g$ fixed and vary $t$. In addition to $\rho$ and $\sigma$, we measure thermal quantities such as average energy $\langle E \rangle$, specific heat per site $C\equiv(\langle E^2 \rangle - \langle E \rangle^2)/{\rm Vol}$, and the third cumulant \cite{Sudbo_C3method}
\begin{equation}
C_3 \equiv \frac{\langle E^3 \rangle-3\langle E \rangle \langle E^2 \rangle + 2\langle E \rangle^3}{{\rm Vol}}, 
\label{C3}
\end{equation}
which can be viewed as a derivative of the specific heat with respect to a temperature-like parameter. 

We will also study the derivative of the superfluid stiffness, $d\rho/dt$, which can be evaluated as 
\begin{equation}
\frac{d\rho(k,t)}{dt}=\left\langle \left|J_x(k)\right|^2 \right\rangle \left\langle E_t \right\rangle -\left\langle \left|J_x(k)\right|^2 E_t \right\rangle.
\end{equation} 
where $E_t$ is the contribution to the energy from the short-range interaction, $E_t=\sum_{r}\vec{J}(r)^2$. 
We study the system in the current loop variables $\vec{J}(r)$ using the directed geometric worm algorithm,\cite{AletSorensen2003_PRE, Prokofiev_1998, Prokofiev_2001, Sandvik1998} modified to keep $\vec{J}_{{\rm tot}}=0$. 

\section{Results}
\begin{figure}
\includegraphics[angle=-90,width=\linewidth]{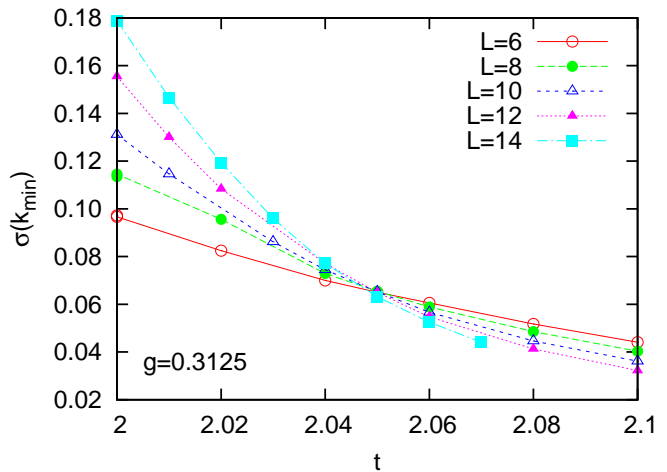}
\caption{An example of the $\sigma(k_{{\rm min}})$ data used to determine the location of the phase transition for the model at fixed $g=0.3125$, varying $t$. From this data, we determine the phase transition to take place at $t_{{\rm crit}}=2.05$, and estimate the crossing value $\sigma^{{\rm crossing}}(k_{{\rm min}})=0.067 \pm 0.005$.}
\label{sigma}
\end{figure}

\begin{figure}
\includegraphics[angle=-90,width=\linewidth]{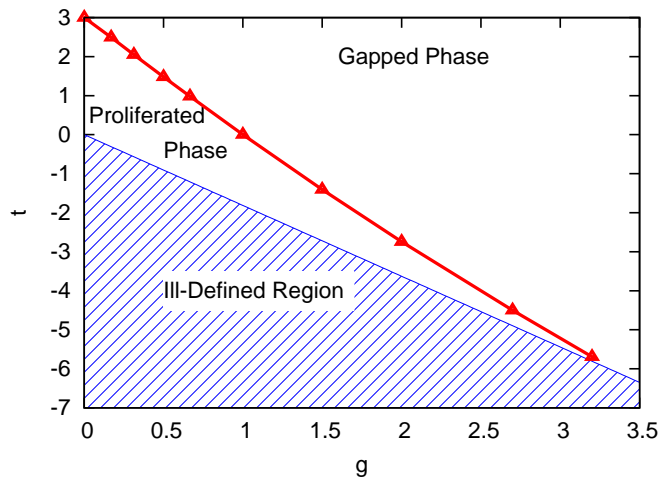}
\caption{The phase diagram of our model, Eq.~(\ref{V}).  The model contains a phase with small loops at large $t$ and a proliferated phase at small $t$, separated by a critical line of continuous phase transitions. The phase boundary is accurately determined from crossings like those in Fig.~\ref{sigma} (see also Table \ref{transition}). For large negative $t$, the model becomes ill-defined. }
\label{phase}
\end{figure}

At $g=1$, the model is exactly self-dual at $t=0$, and symmetry between $\vec{J}$ and $\vec{Q}$ requires that the phase transition between the gapped and proliferated phases occurs here. For $g \neq 1$, we must locate the value of $t$ at which the phase transition occurs. We did this by holding $g$ fixed and sweeping in $t$. In a model with short-range interactions, $\rho \cdot L$ is infinite in the proliferated phase and zero in the small-loops phase, so one could deduce the location of phase transitions from the crossings of $\rho \cdot L$ at different $L$. We saw above that $\rho \cdot L$ does not diverge in the proliferated phase in our long-ranged model, and so we are not guaranteed a crossing. However, we argued that $\sigma$ is zero in the gapped phase and infinite in the proliferated phase in the thermodynamic limit, so we can use the crossings of this quantity to determine the location of the phase transition. An example of this type of crossing is given in Fig.~\ref{sigma}, for $g=0.3125$. The locations of the phase transitions found from these crossings are given in Table~\ref{transition}, and the phase diagram is shown in Fig.~\ref{phase}. 
Note that our model is ill-defined if any of the $V(k)$ become negative. We see from Eq.~(\ref{V}) that this happens first at $k=(\pi,\pi,\pi)$, and $V(\pi,\pi,\pi)<0$ for $t < -g\pi/\sqrt{3}$. This ill-defined region is also labeled in Fig.~\ref{phase}. 

At $g=0$ the model is one of loops with only short-ranged interactions, which is well studied, and our $t_{{\rm crit}}$ agrees with existing results.\cite{Cha1991, AletSorensen2003_PRB} For $g=3.2$, the system was studied in Ref.~\onlinecite{Kuklov2005}, and our $t_{{\rm crit}}$ value agrees with theirs. As a further check on our simulations, it is possible to derive exact values for $\langle E \rangle$ and $\rho \cdot L$ at the self-dual point $g=1$, $t=0$:
\begin{eqnarray*}
\langle E \rangle(g=1,t=0;L) &=& \frac{L^3-1}{2} \\
\rho(g=1,t=0;L) &=& \frac{\sin(\pi/L)}{2\pi}.
\end{eqnarray*}
Our Monte Carlo results are consistent with these relations. 

\begin{table}
\begin{tabular}{|c|ccccccccc|}
\hline
$g$ & 0 & 1/6 & 0.3125 & 0.5 & 2/3 & 1 & 1.5 & 2 & 3.2\\
\hline
$t_{{\rm crit}}(g)$ & 3.01 & 2.50 & 2.05 & 1.48 & 0.98 & 0 & -1.41 & -2.75 & -5.69 \\
\hline
\end{tabular}
\caption{Approximate location of our phase transitions, determined from $\sigma$ crossings; $t_{{\rm crit}}(g)$ are marked in Fig.~\ref{phase}. Error bars are below $0.01$.}
\label{transition}
\end{table} 

\begin{figure}
\includegraphics[angle=-90,width=\linewidth]{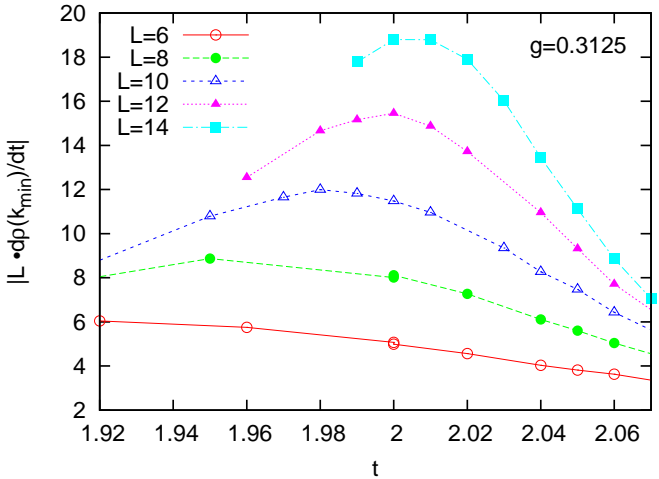}
\caption{$L \cdot d\rho/dt$ as a function of $t$ for the same system as Fig.~\ref{sigma}. Upon increasing $L$, the peak evolves slowly towards the critical $t_{{\rm crit}}=2.05$. For each $L$, we extract the maximum value and show it in Fig.\ref{log}(a). We fit this to Eq.~(\ref{pscale}) to extract $\nu(g)$.}  
\label{peaks}
\end{figure}

\begin{figure}
\includegraphics[angle=-90,width=\linewidth]{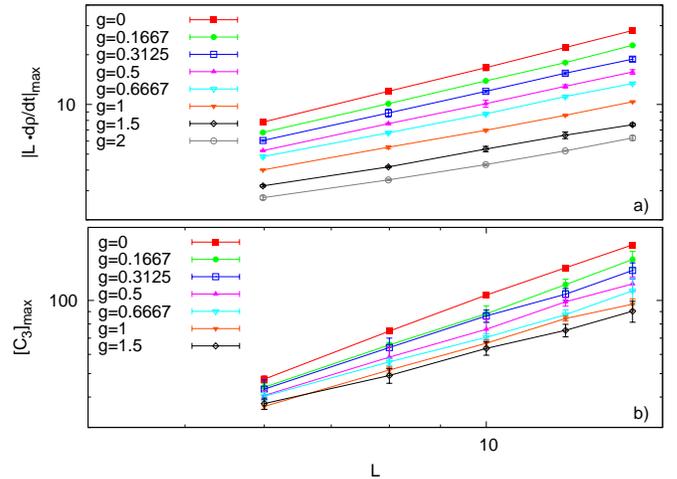}
\caption{Peak values as a function of $L$ for a) $L \cdot d\rho/dt$ and b) $C_3$, plotted on a log-log scale. We can see that the slopes of the lines decrease with increasing $g$ for $0\le g\le1$, which corresponds to an increasing $\nu$. The $C_3$ data at $g=2$ was not shown because it overlaps with the other lines and makes the figure hard to read.}
\label{log}
\end{figure}

\begin{figure}
\includegraphics[angle=-90,width=\linewidth]{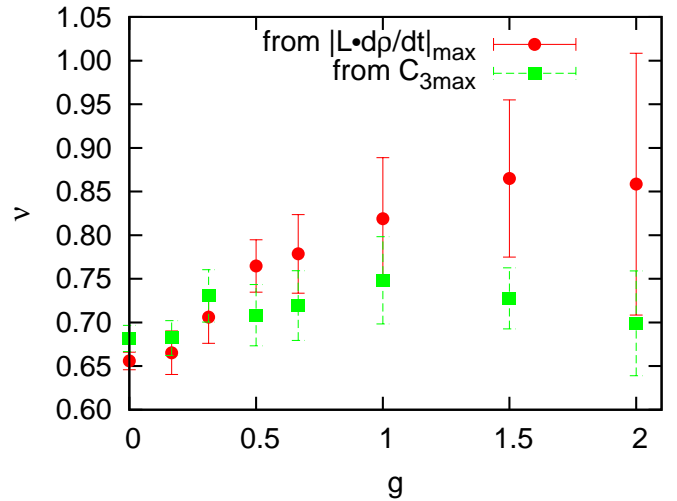}
\caption{Values of $\nu$ determined from scaling of the $|L \cdot d\rho/dt|_{{\rm max}}$ and $[C_3]_{{\rm max}}$, (as shown in Fig.~\ref{log}) with system size. }
\label{nu}
\end{figure}

To study the nature of the phase transition, and in particular the correlation length exponent $\nu$, we looked at the behavior of the derivative $L \cdot \frac{d\rho}{dt}$. Since $\rho \cdot L$ approaches finite values on both sides of the transition, we know that the derivative will be peaked near the critical point. An example of the evolution of $L \cdot \frac{d\rho}{dt}$ is shown in Fig.~\ref{peaks}. Finite-size scaling arguments suggest that $\rho \cdot L=f[(t-t_{{\rm crit}})L^{1/\nu}]$ in our model, and hence the peak value of the derivative behaves as
\begin{equation}
\left[L \cdot \frac{d\rho}{dt}\right]_{ {\rm max}} \sim L^{1/\nu}.
\label{pscale}
\end{equation}
 The extracted peak values as a function of $L$ are shown in Fig. \ref{log}a) for a range of parameters. Fitting Eq.~(\ref{pscale}) gives us values of $\nu$ shown in Fig.~\ref{nu}.  

The same method was applied to $C_3$, which has the peak behavior \cite{Sudbo_C3method}
\begin{equation}
[C_3]_{{\rm max}}\sim L^{3/\nu-d}
\label{Cscale}
\end{equation}
with the dimension $d=3$. We do not show an example of these peaks here, but the extracted $[C_3]_{{\rm max}}$ values are given in Fig.~\ref{log}b) and the extracted values of $\nu$ are shown in Fig.~\ref{nu}. Note that although $C_3$ measures a thermal response while $\rho$ measures a current response, both observables provide similar values for $\nu$. We can see from Fig.~\ref{log} that the curves at large $g$ are not entirely straight on a log-log plot. It was unclear what fitting procedure to use on this data to extract $\nu$, and so we used several different procedures. For small $g$ the lines are nearly straight and $\nu$ was not very dependent on the fitting procedure used, while for larger $g$ we observed a larger change in $\nu$. The error bars in Fig.~\ref{nu} reflect the different values of $\nu$ that can be obtained by changing the fitting procedure. 

We could in principle apply a similar scaling approach to the derivative of the conductivity, $\frac{d\sigma}{dt}$, but we can see from Fig.~\ref{sigma} that these derivatives do not have a maximum whose value we can extract. We could instead extract values at $t_{{\rm crit}}$, but we have found that the results are very sensitive to the estimated value of $t_{{\rm crit}}$, and since we do not in general know $t_{{\rm crit}}$ exactly, we were unable to obtain precise values of $\nu$ in this way. At the self-dual point, where we do know $t_{{\rm crit}}$,  the extracted $\nu$ from the derivatives at the critical point are consistent with the results from the peak values.

We can see that $\nu$ changes as we move along the critical line in Fig.~\ref{phase}. Though our $\nu$ values are not determined very accurately, they are certainly inconsistent with $1/3$, which would be the value suggestive of a first-order transition. \cite{Chala1986} In fact, the values for $g>0$ are all larger than the $\nu$ of the 3DXY model, and so we conclude that our phase transitions are second-order. At $g=0$, the transition is equivalent to the 3DXY model, and our measured value of $\nu$ is consistent with this. 

For the exactly self-dual model, we were able perform simulations at the exact critical point, which allowed us to obtain histograms of energy for sizes up to $L=18$. If the transition were first-order, these histograms would have two peaks.\cite{Chala1986} Our histograms have only one peak, and there is no evidence of a `flat top' which would indicate a double-peak at larger sizes. This further supports our second-order hypothesis. We were also able to estimate $\nu$ using these larger sizes, with results consistent with those reported for $L \le 14$. 

\begin{figure}
\includegraphics[angle=-90,width=\linewidth]{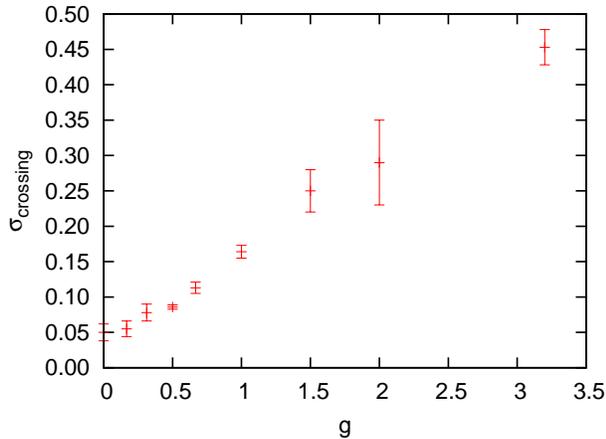}
\caption{The value of $\sigma^{{\rm crossing}}$ vs.~the coupling $g$ of the marginally long-ranged interaction. For each $g$, $\sigma^{{\rm crossing}}$ was determined from plots like those shown in Fig.~\ref{sigma}. We can see that $\sigma^{{\rm crossing}}$ varies as we move along the critical line. The crossings are narrow for $g \le 1$, while for $g>1$, we noticed a weak drift in the crossings as $L$ is increased, and the reported values are from the $L=12$ and $L=14$ data.}
\label{crossings}
\end{figure}

During our study of the phase diagram, we obtained intersections of curves of $\sigma$ for different sizes. The position of these intersections does not noticeably drift with increasing $L$ for $g \le 1$, which further suggests a second-order transition. In Fig.~\ref{crossings} we report the values of these crossings. The reported values are for $L=12$ and $L=14$, since for $g>1$ the crossings did drift slightly with $L$. The error bars for these points are a measure of this drift, while for points at smaller $g$ the crossings at different sizes differ from each other in a non-systematic way (presumably due to statistical errors), and the error bars are a measure of these differences. If $\sigma^{\rm crossing}$ is a universal quantity, we expect it to be determined by only the long-range part of the potential, but for our relatively small sizes it is possible that the short-range term makes a significant contribution. To get an estimate of this, we calculated $\sigma$ without including the short range ($t$) term in the potential in Eq.~(\ref{V}). In the thermodynamic limit, this would not change the result. For our data, the measured value of $\sigma^{{\rm crossing}}$ was lowered by up to $5\%$, with a greater change observed for smaller $g$. This error was in general smaller than that shown in Fig.~\ref{crossings}.

We conjecture that this intersection point $\sigma^{{\rm crossing}}$ is a function of $g$ only. It varies strongly as we move along the line of phase transitions, as shown in Fig.~\ref{crossings}. We also see that $\sigma^{{\rm crossing}}$ evolves continuously into the 3DXY value at $g=0$. From the self-duality of the system at $g=1$, we know from Eq.~(\ref{sigmarelation}) that 
\begin{equation}
\sigma^{{\rm crossing}}(k;g=1)=\frac{1}{2\pi},
\label{specialV}
\end{equation}
which is consistent with our data. It is worth noting that Eq.~(\ref{specialV}) is true for all Matsubara frequencies $i k_z$, and in particular we can analytically continue to real frequencies to obtain the critical dynamical conductivity.\cite{Damle1997, Herzog2007}

\section{Discussion}
We studied a system of loops with $1/r^2$ interactions, and found a line of second-order phase transitions with varying critical properties. We were able to exploit the duality of the model to get some exact results at $g=1$. 

It would be interesting to determine whether the critical properties of the system are dependent solely on the long-range interaction coupling $g$. We could test this by adding additional short-range interactions, e.g.\ a nearest-neighbour interaction, and observing the effect on the critical properties. However, we have seen above that for the accessible sizes, $\sigma$ is sensitive to finite-size effects, and $\nu$ has error bars large enough to obscure any small change due to the new interaction. We could solve both these problems by studying larger sizes.
In these simulations with long-range interactions, increasing the size of the system is quite costly at the present time, but faster computers or parallelization of the energy calculation could make studying these sizes feasible. 

Note that the duality, Eq.~(\ref{duality}), inverts the long-range coupling $g$.\cite{Kuklov2005,Fradkin_SL2Z}  Due to the short-range interaction, the specific model Eq.~(\ref{V}) at $g$ is not exactly dual to the model at $1/g$. However, if $\sigma$ and $\nu$ are dependent only on $g$ and not on the short-range coupling, this would imply 
\begin{eqnarray}
\sigma^{\rm crossing}(1/g) &=& \frac{1}{(2\pi)^2 \sigma^{\rm crossing}(g) }, \label{sigmadual} \\
\nu(1/g) &=& \nu(g). \label{nudual}
\end{eqnarray}
Because $\nu$ varies slowly and has large error bars, we were unable to confirm Eq.~(\ref{nudual}). The relation in Eq.~(\ref{sigmadual}) is satisfied for $0.5 \le g \le 2$, but is not satisfied for the pair $g=0.3125$ and $g=3.2$. However, the crossings at $g=3.2$ seem to be drifting towards a value that would satisfy the relation as $L$ is increased. Studying the system at larger sizes would enable us to determine whether the above relations are satisfied in the thermodynamic limit. We could also study the system at larger $g$, using a modified short-ranged interaction to reduce the size of the ill-defined region in Fig.~\ref{phase}. 

A field-theoretic treatment of the matter-gauge model with the singular gauge propagator is possible in the spirit of Refs.~\onlinecite{HLM1974, Chen1978, ColemanWeinberg1973}.  An analysis suggests a line of fixed points controlled by the coupling $g$ and continuously evolving out of the 3DXY fixed point, and it would be interesting to study this in detail.  It would also be interesting to study multi-component systems realized with multi-loop lattice models\cite{Balents2005, Herzog2007, Kuklov2004, Kuklov2006, Smiseth2005, shortlight, nccp1new, Herland2010} with such marginally-long-ranged interactions. We would also like to study models where loops have mutual statistics and explore the interplay with the marginal interactions.\cite{WenWu1993, ChenFisherWu1993, Fradkin_SL2Z, Sachdev1998}  Loop models with continuously varying critical indices can be a fascinating toolbox for studying phase transitions in three-dimensional statistical mechanics.

{\it Acknowledgments.} We are grateful to M.~P.~A.~Fisher, N.~Read, T.~Senthil, A.~Vishwanath, and W.~Witczak-Krempa for stimulating discussions.   We acknowledge support from the National Science Foundation through grant DMR-0907145; from the Caltech Institute of Quantum Information and Matter, an NSF Physics Frontiers Center with support of the Gordon and Betty Moore Foundation; and from the XSEDE computational initiative grant TG-DMR110052.

\bibliography{bib4rangeloops}
\end{document}